\documentclass[
    twocolumn,
    showpacs, notitlepage, nofootinbib,
    floatfix,superscriptaddress,
    pra,
    longbibliography,
    aps,
]{revtex4-1}

\usepackage[utf8]{inputenc}
\usepackage[english]{babel}
\usepackage[T1]{fontenc}
\usepackage{amsmath,amssymb} 
\usepackage{amsthm}
\usepackage{amsfonts}
\usepackage{euscript}
\usepackage{graphicx}
\usepackage{xcolor}
\usepackage[breaklinks=true,colorlinks=true,pdfusetitle=true]{hyperref}
\usepackage[ruled, vlined, linesnumbered,nofillcomment]{algorithm2e}

\usepackage{tikz}
\usetikzlibrary{patterns}

\usetikzlibrary{decorations.pathreplacing}
\usepackage{stmaryrd} %\rrbracket
\usepackage{url}

\usepackage{braket}

\makeatletter
\newtheorem*{rep@theorem}{\rep@title}
\newcommand{\newreptheorem}[2]{%
\newenvironment{rep#1}[1]{%
 \def\rep@title{#2 \ref{##1}}%
 \begin{rep@theorem}}%
 {\end{rep@theorem}}}
\makeatother

\newtheorem{theo}{Theorem} %[section]
\newreptheorem{theo}{Theorem}

\newtheorem{lemma}[theo]{Lemma}

\newtheorem{corol}[theo]{Corollary}

\usepackage{thmtools}
\usepackage{thm-restate}

\def\L{\mathcal{L}}

\def\tr{\mbox{tr}}

\def\1{\mathbbm{1}}

\DeclareUnicodeCharacter{00A0}{ }

\newcommand{\M}{\mathcal M}
\newcommand{\N}{\mathcal N}

\renewcommand{\geq}{\geqslant}
\renewcommand{\leq}{\leqslant}

\usepackage{bbm}

\begin{document}

\title{Quantum error-correcting codes with a covariant encoding}
\author{Aur\'elie Denys}
\email{aurelie.denys@quandela.com}
\address{Inria Paris, France}  \address{Quandela}
\author{Anthony Leverrier}
\email{anthony.leverrier@inria.fr}
\address{Inria Paris, France}

\date{\today}

\begin{abstract}
Given some group $G$ of logical gates, for instance the Clifford group, what are the quantum encodings for which these logical gates can be implemented by  simple physical operations, described by some physical representation of $G$? We study this question by constructing a general form of such encoding maps. For instance, we recover that the $\llbracket 5,1,3\rrbracket$ and Steane codes admit transversal implementations of the binary tetrahedral and binary octahedral groups, respectively.
For bosonic encodings, we show how to obtain the GKP and cat qudit encodings by considering the appropriate groups, and essentially the simplest physical implementations. We further illustrate this approach by introducing a 2-mode bosonic code defined from a constellation of 48 coherent states, for which all single-qubit Clifford gates correspond to passive Gaussian unitaries. 
\end{abstract}

\maketitle

A main challenge in designing fault-tolerant approaches to quantum computing is the need to address two seemingly conflicting requirements: protecting quantum information against various sources of noise, and manipulating the same quantum information in order to perform a computation. 
The standard approach is to first look for quantum error-correcting codes offering a good protection against relevant noise sources, and then try to understand how to implement logical gates fault-tolerantly, meaning that they shouldn't map innocuous, correctable errors to uncorrectable ones.
Here, we explore the other direction: starting with some logical gate set of interest, we design all the encodings compatible with a given physical implementation of these gates, and only later pick the ones with good error protection capabilities. 

Consider a logical Hilbert space $\mathcal{H}_L$ (typically, a logical qubit) and a physical Hilbert space $\mathcal{H}_P$ (for instance, an $n$-qubit space, or a multimode Fock space for bosonic encodings), as well as two unitary representations $\lambda$ and $\pi$ of some group $G$ on these two Hilbert spaces (See. For each $g \in G$, the operators $\lambda(g)$ and $\pi(g)$ will correspond to unitary operations on the logical and the physical spaces, respectively. What are the isometric encodings $\mathcal{E}: \mathcal{H}_L \to \mathcal{H}_P$ covariant with respect to $G$, \textit{i.e.} such that 
\begin{align}\label{eqn:G-cov}
\pi (g) \, \mathcal{E}(|\psi\rangle) = \mathcal{E}( \lambda(g) \, |\psi\rangle),
\end{align}
for all $g \in G$ and $|\psi\rangle \in \mathcal{H}_L$?
This property means that the physical operation $\pi(g)$ implements the logical operation $\lambda(g)$. Details can be found in App.~\ref{app:iso} of the supplementary material, which includes Ref.~\cite{kna86}.
Ensuring the covariance property is easy: averaging any linear map $V: \mathcal{H}_L \to \mathcal{H}_P$ over the group action gives rise to a $G$-covariant map $V_G$:
\begin{align}\label{eqn:VG}
V_G := \frac{1}{|G|} \sum_{g \in G} \pi(g) V \lambda(g)^\dagger.
\end{align}
How to choose $V$ to get an isometry, and therefore a valid quantum encoding?
Our main technical result (Lemma~\ref{lem:representation} below) provides a sufficient condition. 

We illustrate our approach by first recovering several well-known quantum codes as instances of this construction, and then by introducing a multimode bosonic code with a simple implementation of the single-qubit Clifford group. For multiqubit codes, transversal physical gates are a natural choice of fault-tolerant implementation: $\pi(g) = \lambda(g)^{\otimes n}$. The 5-qubit code and the Steane code are examples of small codes with a transversal implementation of the binary tetrahedral group $2T$ and binary octahedral group $2O$ (\textit{i.e.} the single-qubit Clifford group).
Moving on to bosonic codes~\cite{TCV20,alb22}, we explain how the GKP code~\cite{GKP01} and cat codes~\cite{CMM99,MLA14} fit into the formalism. These two constructions are respectively obtained for a physical representation $\pi$ generated by linear or quadratic Hamiltonians, that is, displacements or phase-shifts. 
We also design a multimode bosonic code for which the single-qubit Clifford group is implemented with passive Gaussian unitaries (phase-shifts and beamsplitters). We further show how to obtain a universal gate set by adding quartic Hamiltonians. The advantage of this code compared to the GKP code is that its encoded states are superpositions of a finite number of coherent states, removing any normalization issue.

We note that this construction is not limited to multiqubit or bosonic encodings. In fact, Gross applied a similar approach to design codes with covariant gates on spin systems~\cite{gro21}.\\

{\bf A sufficient condition for isometric encodings}.--- For finite groups $G$, the following lemma provides a simple condition for the map $V_G$ to be isometric.

\begin{lemma}\label{lem:representation}
Let $V$ be a linear map between a logical Hilbert space $\mathcal{H}_L \cong \mathbb{C}^d$ and a physical Hilbert space $\mathcal{H}_P$. Let $G$ be a finite group with two unitary representations $\lambda, \pi$ on $\mathcal{H}_L$ and $\mathcal{H}_P$, respectively. Assume in addition that $\lambda$ is irreducible. If $v := \tr(V^\dag V_G)/d \ne 0$, then $v^{-1/2} V_G$ is a $G$-covariant isometry.
\end{lemma}

The proof is deferred to App.~\ref{app:prelim} and follows from the Schur orthogonality relations.
The lemma extends to compact topological groups such as $SU(d)$ thanks to the Peter-Weyl theorem. In this case, the averaging in \eqref{eqn:VG} is with respect with the Haar measure over $G$.
 The assumption that $\lambda$ is irreducible is sufficient but not necessary to obtain an isometric encoding. We discuss the example of the  bosonic cat and rotation-symmetric codes~\cite{GCB20} in App.~\ref{app:general}.

An easy application of the lemma concerns rank-1 operators $V$ of the form $|\Phi\rangle \langle \Omega|$, where $|\Omega\rangle \in \mathcal{H}_L$ and $|\Phi\rangle\in \mathcal{H}_P$. In particular, \eqref{eqn:VG} implies that the codewords are superpositions of states of the form $\pi(g) |\Phi\rangle$. In the case of bosonic codes for instance, choosing $|\Phi\rangle$ to be a coherent state and $\pi(g)$ to be a passive Gaussian unitary ensures that the codewords are superpositions of coherent states.\\

{\bf Encoding circuit for $V = |\Phi\rangle \langle \Omega|$}.--- Under the assumptions of Lemma \ref{lem:representation}, it is possible to write down an explicit circuit for the encoding map. Let $U$ be a unitary operator such that the physical representation decomposes as $\pi(g) = U( \lambda(g) \otimes \1_{\M} \oplus \pi'(g))U^\dagger$ where $\M$ is the multiplicity space associated to $\lambda$, and $\lambda$ is not contained in the remaining representation $\pi'$.  
The encoding map can be understood as follows: the state $|\Phi\rangle$ is first projected onto the isotypic component of $\lambda$, then  $U^\dag$ maps it to a bipartite state $U^\dag \Pi |\Phi\rangle \in \mathcal{H}_{L'} \otimes \M$, where $\mathcal{H}_{L'} \cong \mathcal{H}_L$. The registers $\mathcal{H}_L$ and $\mathcal{H}_{L'}$ are swapped. The first system is projected onto $|\Omega\rangle$ while the remaining logical state is embedded back into $\mathcal{H}_P$ through the application of $U$.
The final encoding only depends on the choice of the specific state $|\phi\rangle \propto \langle \Omega | U^\dag |\Phi\rangle \in M$, and that the encoding map takes the simplified form
\begin{align}\label{eqn:alt-enc}
 \mathcal{E}(|\psi\rangle) = U |\psi\rangle |\phi\rangle.
\end{align}
This is displayed on Fig.~\ref{fig1} and detailed in App.~\ref{app:circuit}.

\begin{figure}
\scalebox{0.75}{
\begin{tikzpicture}

\coordinate (B1) at (0,-3); \coordinate (B2) at (1,-3);
\coordinate (C) at (1.5,-3); \coordinate (C1) at (1,-3+0.5); \coordinate (C4) at (1,-3-0.5); \coordinate (C2) at (2,-3+0.5); \coordinate (C3) at (2,-3-0.5);
\coordinate (B3) at (2,-3); \coordinate (B4) at (3,-3);
\coordinate (D) at (3.5,-3); \coordinate (D1) at (3,-3+0.5); \coordinate (D4) at (3,-3-0.5); \coordinate (D2) at (4,-3+0.5); \coordinate (D3) at (4,-3-0.5); \coordinate (D5) at (4.5,-3+0.15); \coordinate (D6) at (4.75,-3-0.15);
\coordinate (E1) at (4,-3+0.15); \coordinate (E2) at (5,-3+0.15); \coordinate (E3) at (6,-1.12); \coordinate (E4) at (8,-1.12); \coordinate (E5) at (8.5,-1.12);
\coordinate (F1) at (4,-3-0.15); \coordinate (F2) at (7,-3-0.15);
\coordinate (A1) at (0,-1.12); \coordinate (A2) at (5,-1.12); \coordinate (A3) at (6,-3+0.15); \coordinate (A4) at (7,-3+0.15);
\coordinate (F) at (7.5,-3); \coordinate (F1) at (7,-3+0.5); \coordinate (F4) at (7,-3-0.5); \coordinate (F2) at (8,-3+0.5); \coordinate (F3) at (8,-3-0.5); \coordinate (F5) at (8.5,-3);
\coordinate (B5) at (8,-3); \coordinate (B6) at (8.5,-3);
\coordinate (G1) at (4, -3-0.15); \coordinate (G2) at (7, -3-0.15);

\draw (A1)--(A2)--(A3)--(A4);  
\draw[opacity=0.6] (B1)--(B2); \draw[opacity=0.6] (B3)--(B4); \draw[opacity=0.6] (B5)--(B6);  
\draw[opacity=0.6] (C1)--(C2)--(C3)--(C4)--cycle;  
\draw[opacity=0.6] (D1)--(D2)--(D3)--(D4)--cycle;  
\draw[opacity=0.6] (E1)--(E2)--(E3)--(E5);  
\draw (F1)--(F2)--(F3)--(F4)--cycle;  
\draw (G1)--(G2);  

 \draw[opacity=0.6] (B1) node[left] {$| \Phi \rangle \in \mathcal{H}_P $} ;
\draw[opacity=0.6] (E5) node[right] {$\langle \Omega |$} ; 
\draw (A1) node[left] {$| \psi \rangle \in \mathcal{H}_L$} ;

\draw[opacity=0.6] (C) node {$\Pi$} ;
\draw[opacity=0.6] (D) node {$U^\dagger$} ;
\draw (F) node {$U$} ;
\draw (F5) node[right] {$\mathcal{E}(|\psi\rangle)$} ;

\draw[opacity=0.6] (D5) node[above] {$\mathcal{H}_{L'}$} ;
\draw (D6) node[below] {$\M \ni |\phi\rangle$} ;

\end{tikzpicture}
}
\caption{Encoding circuit of Lemma \ref{lem:representation}, when $V = |\Phi\rangle \langle \Omega|$. The unitary operator $U$ block-diagonalizes the physical representation $\pi$ and $\Pi$ projects onto the irreducible representation of $\lambda$ in $\mathcal{H}_P$. After the gate $U^\dag$, the state lives in $\mathcal{H}_L\otimes \mathcal{H}_{L'} \otimes \M$, with $\M$ the multiplicity space of $\lambda$. Registers $\mathcal{H}_{L}$ and $\mathcal{H}_{L'}$ are swapped. The first register is projected onto the state $|\Omega\rangle$, while the remaining registers are embedded back to the physical space thanks to $U$. The role of the circuit depicted in gray is to prepare a state $|\phi\rangle \in \M$. Ignoring that part, one recovers \eqref{eqn:alt-enc}. } 
\label{fig1} 
\end{figure}

$G$-covariant encodings can thus be specified, either through a couple $(|\Omega\rangle, |\Phi\rangle) \in \mathcal{H}_L \times \mathcal{H}_P$, or through a single state $|\phi\rangle \in \M$. This choice influences the correction capabilities of the code, either its distance for a multiqubit code, or its performance against a given noise channel for more general codes. Both encoding forms can be useful: in some cases, there is a natural family of physical states $|\Phi\rangle$ that can be optimized over, \textit{e.g.}\ coherent states for bosonic codes. In other cases, it is more efficient to directly exhaustively search over the multiplicity space $\M$. \\

We note that when the logical representation $\lambda$ is irreducible, then our construction is general in the sense that any encoding $\mathcal{E}$ satisfying \eqref{eqn:G-cov} is proportional to $V_G$, with $V := \mathcal{E}(|0\rangle)\langle 0|$.
We now show how to recover well-known codes with this construction, starting with multiqubit codes and then moving on to bosonic encodings.

{\bf Small codes with large transversal gate sets}.---
Transversal gates are an attractive choice of physical gates: they can be applied in depth one and are therefore fault tolerant. Exactly transversal gates satisfy $\pi(g) = \lambda(g)^{\otimes n}$. 
The Eastin-Knill theorem puts severe restrictions on the set of transversal gates for a nontrivial quantum code~\cite{EK09,FNA20}, suggesting some trade-off between the size of the group $G$ and the error protection offered by the encoding. The recent works by Kubischta and Teixeira~\cite{KT23,KT23b,KT24} study this interplay by exhibiting nontrivial codes with exotic sets of transversal gates. 
Here, we focus on the simplest cases corresponding to the binary tetrahedral group $2T$ and the binary octahedral group $2O$, which are finite subgroups of respective order $|2T|=24$ and $|2O|=48$ of $SU(2)$, given by 
\begin{align}
2T = \langle \mathrm{Z}, \mathrm{H}\rangle, \qquad 2O = \langle \mathrm{H}, \mathrm{S}\rangle
\end{align}
with the specific choice of matrices
\begin{align}
\mathrm{Z}= \left[ \begin{smallmatrix} i & 0 \\ 0 & -i\end{smallmatrix}\right], \quad \mathrm{S} = \left[ \begin{smallmatrix} \eta & 0 \\ 0 & \eta^{-1}\end{smallmatrix} \right], \quad \mathrm{H} = \frac{1}{\sqrt{2}} \left[ \begin{smallmatrix} \eta & \eta \\ -\eta^{-1} & \eta^{-1}\end{smallmatrix} \right]
\end{align}
with $\eta = e^{i \pi/4}$. This choice is similar to~\cite{KT23} and guarantees that $\mathrm{Z}, \mathrm{S}, \mathrm{H} \in SU(2)$.

Applying the construction of Lemma \ref{lem:representation} to these groups yield $n$-qubit encodings with transversal gates, and the question is whether one can find an auxiliary system $|\phi\rangle \in M$ such that the resulting code given by \eqref{eqn:alt-enc} admits a nontrivial distance.
Eastin-Knill states that any state $|\phi\rangle$ in the multiplicity space of the irreducible representation of $SU(2)$ gives a trivial code. Therefore, one needs to pick a value of $n$ such that the dimension of the multiplicity space associated with $2T$ or $2O$ is strictly larger than that of $SU(2)$. 
This occurs for $n=5$ for the group $2T$ (the multiplicity spaces associated with $SU(2)$ and $2T$ have dimension 5 and 6, respectively) and $n=7$ for the group $2O$ (the dimensions of the multiplicity spaces are 14 and 15).
In both cases, the suitable choice of $|\phi\rangle$ in the multiplicity space yields the $\llbracket 5,1,3\rrbracket$ code and a code equivalent to the Steane $\llbracket 7,1,3\rrbracket$ code, which admit exactly transversal implementations of $2T$ and $2O$ respectively\footnote{The standard definition of the Steane code does not admit a strongly transversal implementation of the Clifford group, but this can be fixed by a simple change of basis: $\mathcal{E}(|0\rangle) := -i |1\rangle_{\text{Steane}}, \mathcal{E}(|1\rangle) :=  |0\rangle_{\text{Steane}}$.}. 
Interestingly, these codes achieve the minimum possible number of physical qubits allowed for nontrivial codes with these sets of transversal gates. These observations were already made in \cite{KT23}.

While our construction can recover these codes, it is not clear at the moment how to efficiently find the best possible choice of $|\phi\rangle$ in the multiplicity space that yields codes with a large distance. An approach in this direction is laid out in \cite{KT24}, but only seems applicable to very small distances. We show next that for bosonic codes, there is often a natural choice of state $|\Phi\rangle$ leading to interesting encodings. \\

{\bf The GKP code}.--- The GKP code is one of the most studied bosonic codes, with excellent performance against random displacements in phase space, as well as pure loss~\cite{GKP01, AND18, GP21}. It slightly deviates from the construction of Lemma \ref{lem:representation} because the relevant group, the discrete Heisenberg group, is infinite, thus leading to unnormalizable states. Ignoring this issue here, let us see how to recover the GKP encoding.  The discrete Heisenberg group is $G = H_3(\mathbb{Z}) = \langle x,y \rangle$ with $x= \left[\begin{smallmatrix} 1 & 1 &0 \\ 0 & 1 & 0 \\ 0 & 0 & 1 \end{smallmatrix}\right], \qquad y = \left[\begin{smallmatrix} 1 & 0 & 0 \\ 0 & 1 & 1 \\ 0 & 0 & 1\end{smallmatrix}\right].$ We consider two representations of the Heisenberg group. 
The $d$-dimensional logical representation is given by
\begin{align}
\lambda(x) = X, \qquad \lambda(y) = Z,
\end{align}
where $X$ and $Z$ are the generalized $d$-dimensional Pauli matrices.
The physical representation $\pi$ is infinite-dimensional and acts as displacements on a single bosonic mode with creation and annihilation operators $\hat{a}^\dagger$ and $\hat{a}$:
\begin{align}
\pi(x) = \hat{D}(\alpha), \qquad \pi(y) = \hat{D}(\beta),
\end{align}
where $\alpha, \beta \in \mathbbm{C}$ are chosen to satisfy $e^{\beta \alpha^* - \beta^* \alpha} =e^{2\pi i/d}$ and the displacement operators are defined by $\hat{D}(\alpha) = e^{\alpha \hat{a}^\dagger - \alpha^* \hat{a}}$.
Once the group $G$ and these reasonably natural representations are fixed, the only remaining freedom is the choice of the physical and logical states $|\Phi\rangle$ and $|\Omega\rangle$. Taking the simplest possibility, namely the vacuum state for $|\Phi\rangle$ and $|\Omega\rangle \propto |0\rangle$ yields the standard GKP encoding:
\begin{align}
\mathcal{E}(|k\rangle) \propto  \sum_{a,b \in \mathbbm{Z}} (e^{2\pi i/d})^{abd/2 - bk/2} | (da + k) \alpha + b \beta \rangle.
\end{align}
In particular, $\alpha = \sqrt{\pi/2}$, $\beta = i \alpha$ with $d=2$ give the GKP qubit with a square lattice. 
See App.~\ref{app:gkp}.\\

{\bf The cat qudit code}.--- 
Cat encodings are increasingly popular bosonic codes that may offer a dramatic hardware reduction for quantum fault tolerance~\cite{GM19,PSG20,RGL24}.
They are obtained from constellations of coherent states of the form $\{ |\omega^p \alpha\rangle\}_{0\leq p <n}$ for some $\alpha>0$, and $\omega = e^{2\pi i/n}$. We choose $n = md$ to encode a $d$-dimensional qudit with $n$ coherent states. These codes have the property that physical operations are obtained by phase-shifts of the form $\omega^{p \hat{n}}$.
They illustrate our framework with the cyclic group $G = C_n := \{ \omega^p \: : \: 0\leq p \leq n-1\}$ and its representations
\begin{align}
\lambda(\omega^p) = Z^p \quad \text{and} \quad \pi(\omega^p) = \omega^{p \hat{n}}.
\end{align}
Lemma \ref{lem:representation} doesn't directly apply here since the logical representation $\lambda$ is not irreducible: it decomposes as a sum of $d$ irreducible representations, all with trivial multiplicity: for all $g \in G$, $\lambda(g) =  \bigoplus_{k=0}^{d-1} \rho_{k}(g)$, where $\rho_k(\omega^p) = \omega^{kp}$. In particular, not all choices of the state $|\Omega\rangle \in \mathbbm{C}^d$ give rise to an isometric encoding. Rather, one must choose the corresponding state with some care. We prove in App.~\ref{app:general} that one can still obtain an isometric encoding for any $|\Phi\rangle$ when all the multiplicity spaces of $\lambda$ are trivial. One then recovers the cat qudit encoding by setting $|\Phi\rangle$ to be a coherent state $|\alpha\rangle$.\\

{\bf Multimode bosonic codes}.--- 
Beyond recovering known codes, we can construct new bosonic codes encoding a qubit: $\mathcal{H}_L = \mathbbm{C}^2$. We will pick two subgroups of $U(2)$: the Pauli group and the single-qubit Clifford group $2O$ and their standard 2-dimensional irreducible representations.  We choose $\mathcal{H}_P$ to be a two-mode Fock space with annihilation operators $\hat{a}_1, \hat{a}_2$. The natural representation $\pi$ we consider maps the unitary $U \in U(2)$ to the passive Gaussian unitary acting on the creation operators as follows~\cite{WPG12}: $\pi(U): (\hat{a}_1^\dag, \hat{a}_2^\dag) \mapsto (\hat{a}_1^\dag, \hat{a}_2^\dag) U$, and mapping a coherent state $|\vec{\alpha}\rangle$ to $|U \vec{\alpha}\rangle$. 
Finally, we choose for $|\Phi\rangle\in \mathcal{H}_P$ a two-mode coherent state $|\alpha\rangle|\beta\rangle$. Because $\pi(g) |\alpha\rangle |\beta\rangle$ is again a coherent state, the codewords are superpositions of $|G|$ coherent states. 
The resulting codes are reminiscent of quantum spherical codes~\cite{JIB23} in the sense that the coherent states appearing in the constellations all lie on a sphere in phase-space.  \\

{\bf The bosonic Pauli code}.--- Several versions of the Pauli group exist, e.g, $\langle X, Z\rangle$ of order 8 and $\langle i, X, Z\rangle$ of order 16. Picking $\langle X, Z\rangle$ and $|\Omega\rangle \propto |0\rangle$, the construction discussed above gives
\begin{align}\label{eqn:pauli}
\mathcal{E}(|0\rangle) \propto |c_1(\alpha)\rangle |c_0(\beta)\rangle, \quad \mathcal{E}(|1\rangle) \propto |c_0(\beta)\rangle|c_1(\alpha)\rangle,
\end{align}
where  $|c_k(\alpha)\rangle  := |\alpha\rangle + (-1)^k |-\alpha\rangle$ denotes an unnormalized single-mode cat state.
One recovers the dual-rail encoding~\cite{KLM01,CSM23,LHH23} in the limit $\alpha, \beta \to 0$. Interestingly, the dual-rail code is also an example of the construction for the compact group $G=SU(2)$: while all single-qubit gates can be implemented with passive Gaussian unitaries, it is a rather bad code since it can only detect a single photon loss, but cannot correct it.

By construction, logical Pauli operators can be implemented with Gaussian unitaries on the codewords of \eqref{eqn:pauli}: swapping the modes gives a logical $X$-gate and the phase-gate $(-1)^{\hat{n}_2}$ on the second mode implements a logical $Z$-gate.
We detail in App.~\ref{app:pauli} how to obtain a logical $S$ gate $\left[\begin{smallmatrix} 1 & 0 \\ 0 & i\end{smallmatrix}\right]$ and a $CZ$ gate by applying quartic Hamiltonians, corresponding respectively to unitaries $i^{\hat{n}_2^2}$ and $(-1)^{\hat{n}_2\otimes \hat{n}_4}$, similarly to the case of rotation-symmetric bosonic codes~\cite{GCB20}.

\begin{figure}
\includegraphics[width=\linewidth]{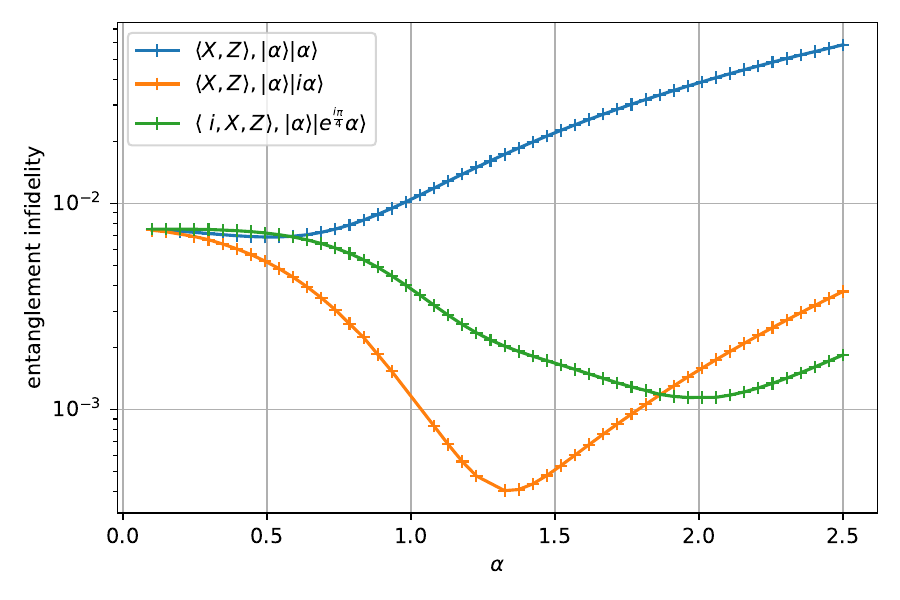} 
\caption{Entanglement infidelity for the pure-loss channel with loss rate $\gamma=10^{-2}$ for 3 variants of the Pauli code, depending on the group $\langle X, Z\rangle$ or $\langle i , X, Z\rangle$ and initial state.  The value at $\alpha=0$ corresponds to the dual-rail encoding.}
\label{fig}
\end{figure}

We simulate the performance for a pure-loss channel, as in~\cite{AND18} and~\cite{DL23}, and plot on Fig.~\ref{fig} the entanglement infidelity assuming optimal recovery. This figure of merit has the advantage of being efficiently computable when the constellation size is not too large. It provides some insight about the protection offered by the encoding. If the group is $\langle X, Z\rangle$,  then the initial state $|\alpha\rangle |\alpha\rangle$ yields a constellation of minimal size, but the choice $|\alpha\rangle|i\alpha\rangle$ provides a much better tolerance to loss, also compared to the dual-rail encoding. This is consistent with an analysis of the Knill-Laflamme conditions for the Kraus operators of the pure-loss channel, as discussed in App.~\ref{app:KL}. 
The variant $\langle i, X, Z\rangle$ of the Pauli group leads to a larger constellation, but doesn't improve the protection against loss, even for an optimized choice of initial coherent states satisfying $\beta = e^{i\pi/4}\alpha$. \\

{\bf The bosonic Clifford code}.---  We now apply the construction to the single-qubit Clifford group $2O$. Again, all the gates from this group are implemented with beamsplitters and phase-shifters. We show in App.~\ref{app:proof3} how to implement the $CZ$ and $T$ gates with quartic Hamiltonians, providing a universal gate set. 
Choosing as before the initial state $|\Phi\rangle =|\alpha\rangle |\beta\rangle$ and $|\Omega\rangle \propto |0\rangle$ gives logical states $\mathcal{E}(|0\rangle)$ and $\mathcal{E}(|1\rangle)$ which are superpositions of 40 coherent states (see App.~\ref{app:cliff}).\\

{\bf State preparation and measurements}.--- The codes obtained through Lemma \ref{lem:representation} are not stabilizer codes in general, which suggests that state preparation may be complex. 
For the Pauli code of \eqref{eqn:pauli}, however, the two logical basis states are product states with a cat state in each mode. Such states are routinely prepared and manipulated in the lab today, and we expect state preparation to be feasible in the near term. 
On the other hand, the preparation of a logical state of the Clifford code seems significantly more delicate.

Measuring the states in the (logical) computational basis can be done by measuring the photon number in each mode. This is because $\mathcal{E}(|0\rangle)$ is a superposition of Fock states of the form $|\text{odd}\rangle |\text{even}\rangle$ while the states in $\mathcal{E}(|1\rangle)$ are of the form $|\text{even}\rangle|\text{odd}\rangle$ for the Pauli code, and $\mathcal{E}(|k\rangle)$ is a superposition of Fock states of the form $|8m+p+1\rangle |8n + p +2k\rangle$ for the Clifford code. Techniques developed for cat codes may prove useful here.\\

{\bf Discussion and open questions}.---
We have introduced a general methodology for designing quantum error-correcting codes that admit a specific logical group implementable with simple physical gates, either transversal gates for qubit codes or Gaussian unitaries for bosonic codes. 
In the latter case, one can design a code with a universal gate set consisting of such Gaussian unitaries together with some gates corresponding to quartic Hamiltonians. While these gates are certainly more challenging to implement, they do not seem out of reach for circuit QED~\cite{BGG21}. On the other hand, such gates are much more difficult for photonic implementations, and it would be interesting to understand whether other gadgets can be used to obtain a universal gate set in that case. 
A question that we have not addressed at all is how to perform error correction. More generally, the stabilization of such states appears quite daunting since they involve rather large constellations of coherent states, but this might not necessarily be more complicated than stabilizing a GKP state~\cite{SSL23}. 

Beyond bosonic codes, it will be interesting to apply this formalism to other physical systems, for instance rotors~\cite{VCT24} or molecules~\cite{ACP20}.
Concerning multiqubit codes, we have shown that it is easy to recover known codes, basically by picking the right state $|\phi\rangle$ in the multiplicity space. What is unclear is whether there is a simple way to find such a state, when it is not known in advance.

\medskip

\acknowledgements{We thank Mazyar Mirrahimi and Christophe Vuillot for many discussions about bosonic codes. We also thank Victor Albert and Jonathan Gross for discussions on representation theory, Simon Burton for discussions on the relation between the Clifford group and the binary octahedral group, Markus Heinrich, Richard K\"ung and Dominik Wild for answering a question about unitary $t$-designs, and Eric Kubischta and Ian Teixeira for discussions about transversal gates for multiqubit codes. We acknowledge support from the Plan France 2030 through the project NISQ2LSQ ANR-22-PETQ-0006.}

%bibliography{biblio}


\newpage
\appendix
\onecolumngrid

\section*{Supplementary material}

We start this supplementary material by defining isometric covariant encodings in Section~\ref{app:iso} before giving brief preliminaries on the representation theory of finite groups in Section \ref{app:prelim}. Sections \ref{app:proof} and \ref{app:circuit} then provide the proof of Lemma \ref{lem:representation} in the main text as well as the corresponding circuit description of the encoding map. Section \ref{app:gkp} provides details on the GKP construction. Section \ref{app:general} is devoted to a relaxation of the assumptions of Lemma \ref{lem:representation}. In particular, Corollary \ref{corol} illustrates the case where the logical representation $\lambda$ is not irreducible, but contains several irreducible representations with multiplicity 1. This is the setup relevant to recover the cat qudit encoding. Section \ref{app:gates} presents possible implementations of physical gates that are not included in the group $G$, for the multimode bosonic encodings introduced in the main text. Section \ref{app:KL} discusses the Knill-Laflamme conditions for a pure-loss channel applied to these multimode bosonic codes. Finally Section \ref{app:cliff} provides the explicit description of the encoding of the multimode bosonic code obtained when the group $G$ is the single-qubit Clifford group.

\section{Isometric covariant encodings}
\label{app:iso}
In this manuscript, we are concerned with \emph{covariant} encodings, which are linear maps $\mathcal{E}$ from a logical space $\mathcal{H}_L$ to a physical space $\mathcal{H}_P$ that commute with a group action in the following sense: given some group $G$ and two representations $\lambda$ and $\rho$ of this group on $\mathcal{H}_L$ and $\mathcal{H}_P$, respectively, we say that $\mathcal{E}$ is covariant with these group representations if
\begin{align}
\pi(g) \circ \mathcal{E} = \mathcal{E} \circ \mathcal{\lambda}(g) \qquad \forall g \in G.
\end{align}
The maps we have in mind correspond to encodings of quantum error-correcting codes, and as such they should be linear \emph{isometries} from $\mathcal{H}_L$ to $\mathcal{H}_P$. This means that there should exist a linear operator $W \in \mathcal{L}(\mathcal{H}_L, \mathcal{H}_P)$ such that   
\begin{align}
\forall |\psi\rangle \in \mathcal{H}_L, \qquad \mathcal{E}(|\psi\rangle) = W |\psi\rangle \quad \text{and} \quad W^\dagger W = \mathbbm{1}_{\mathcal{H}_L}.
\end{align}
As pointed out in the main text, obtaining a covariant map from an arbitrary operator $V \in \mathcal{L}(\mathcal{H}_L, \mathcal{H}_P)$ is straightforward, simply by averaging it over the group action to get $V_G = \frac{1}{|G|} \sum_{g\in G} \pi(g) V \lambda(g)^\dagger$. The nontrivial part is ensuring that this map is indeed an isometry.

\section{Preliminaries on the representation theory of finite groups}
\label{app:prelim}

Consider some unitary representation $\rho$ of a finite group $G$ on some Hilbert space $\mathcal{H}$. This representation can be decomposed as a direct sum of irreducible representations: for all $g \in G$,
\begin{align}\label{eqn:decomposition}
\rho(g) = U \left( \bigoplus_i \rho_i (g)_{\L_i} \otimes \mathbbm{1}_{\M_i}\right) U^\dagger,
\end{align}
where $U$ is called the intertwiner operator, the index $i$ labels the various irreducible representations $\rho_i$ acting on the space $\L_i$ and $\M_i$ are their respective multiplicity spaces. This means that the representation $\rho$ acts in a block-diagonal fashion on the space $\mathcal{H}$, and that each block decomposes as a product of two spaces $\mathcal{L}_i \otimes \mathcal{M}_i$, where the representation acts as an irreducible representation $\rho_i$ on $\mathcal{L}_i$, and trivially on the multiplicity space $\mathcal{M}_i$.
The existence of an intertwiner means that the representations $\rho$ and $\bigoplus_i \rho_i \otimes \1_{\M_i}$ are equivalent, and therefore that the Hilbert space $\mathcal{H}$ decomposes as 
\begin{align}
\mathcal{H} \cong \bigoplus_i \L_i \otimes \M_i.
\end{align}

The following lemma shows that the intertwiner can be assumed unitary, when the representations are unitary.

\begin{lemma}\label{lem:U}
Let $\pi$ and $\sigma$ be two equivalent unitary representations of a group $G$. Then there exists a unitary intertwiner $U$ such that
\[ U \pi(g) = \sigma(g) U, \qquad \forall g\in G.\]
\end{lemma}
\begin{proof}
Let $T$ be an intertwiner for the two representations:
\[ T \pi(g) = \sigma(g) T\qquad \forall g\in G.\]
Since $\pi(g)$ and $\sigma(g)$ are unitary, we get
\[ T^{\dag (-1)} \pi(g) = \sigma(g) T^{\dag (-1)}.\]
 Define $|T| = \sqrt{ T^\dag T}$. The operator $U = T |T|^{-1}$ is unitary since
 \begin{align*}
 U U^\dag &= T |T|^{-2} T^\dag
= T (T^\dag T)^{-1} T^\dag
 = T  T^{-1} T^{\dag (-1)} T^\dag= \1,\\
 U^\dag U &= |T|^{-1} T^\dag T |T|^{-1}= \1.
 \end{align*}
 Note that $T^\dag T$ commutes with $\pi(g)$:
 \begin{align*}
 T^\dag T \pi(g) &= T^\dag \sigma(g) T = (\sigma(g^{-1}) T)^\dag T
 = (T \pi(g^{-1}))^\dag T
 = \pi(g) T^\dag T.
 \end{align*}
 This implies that all polynomials in $T^\dag T$ also commute with $\pi(g)$, and also all continuous functions, for instance the square-root function. Hence
 \[ |T| \pi(g) = \pi(g)|T|.\]
 Finally
\begin{align*}
U \pi(g) &= T |T|^{-1} \pi(g)= T \pi(g) |T|^{-1}= \sigma(g) T |T|^{-1}= \sigma(g) U,
\end{align*}
which is what we wanted.
\end{proof}

Next, the Schur orthogonality relations show that irreducible representations are orthogonal in the following sense: let $\rho_i$ and $\rho_j$ be two irreducible representations of the finite group $G$, of respective dimension $d_i$ and $d_j$. Then,
\begin{align}
\frac{1}{|G|} \sum_{g \in G} \rho_i(g)^\dagger \otimes \rho_j(g) = 
\left\{ 
    \begin{array}{ll}
0  &\mbox{ if } i \ne j,\\
\frac{1}{d_i} \sum_{p,q=0}^{d_i-1} |p\rangle \langle q| \otimes |q\rangle \langle p|   &\mbox{ if } i=j.
\end{array}
\right.
\end{align}

A last concept that will be useful is the projection of some representation $\rho$ onto the isotypic component associated with some irreducible representation $\rho_i$:
\begin{align}
\Pi_i^\rho := \frac{\mathrm{dim} (\rho_i)}{|G|} \sum_{g\in G} \tr(\rho_i(g))^* \, \rho(g).
\end{align} 
In other words, $\Pi_i^{\rho}$ is the orthogonal projection onto $U (\L_i \otimes \M_i) U^\dagger$.

\section{Proof of Lemma \ref{lem:representation} in the main text}
\label{app:proof}

\begin{proof}[Proof of Lemma \ref{lem:representation}]
Let $V : \mathcal{H}_L \to \mathcal{H}_P$. 
We directly compute $V_G^\dagger V_G$:
\begin{align}
 V_G^\dagger V_G &=  \frac{1}{|G|^2} \sum_{g,h \in G} \lambda(g) V^\dagger  \pi(g)^\dagger \pi(h) V\lambda(h)^\dagger\\
&=\frac{1}{|G|^2}  \sum_{g,h \in G} \lambda(g)V^\dagger \pi(h) V \lambda(h)^\dagger \lambda(g)^\dagger \label{eqn:a6}\\
&=  \frac{1}{d |G|} \sum_{h \in G} \sum_{p,q=0}^{d-1} (|p\rangle \langle q|)_{\mathcal{H}_L}   V^\dagger \pi(h) V\lambda(h)^\dagger (|q\rangle   \langle p|)_{\mathcal{H}_L} \label{eqn:a7} \\
&=  \frac{1}{d |G|} \sum_{h \in G}   \tr(V^\dagger \pi(h) V\lambda(h)^\dagger) \, \1_{\mathcal{H}_L}\\
&= \frac{\tr( V^\dagger V_G)}{d} \, \1_{\mathcal{H}_L}
\end{align}
where \eqref{eqn:a6} is a simple change of variable, \eqref{eqn:a7} is an application of the Schur orthogonality relations for the irreducible representation $\lambda$. In particular, $v = \tr(V^\dagger V_G)/d = \tr(V_G^\dagger V_G)/d \geq 0$. Whenever $v \ne 0$, the map $v^{-1/2} V_G$ is an isometry.
\end{proof}

In the case where the group $G$ is a compact topological group, like $SU(d)$, one needs to modify the averaging corresponding to $V_G$ in the following way:
\begin{align}
V_G &:= \int_{g \in G} \pi(g) V \lambda(g)^\dagger d \mu(g),
\end{align}
where $\mu(g)$ is the Haar measure on the group $G$. The Peter-Weyl theorem~\cite[p.~17]{kna86} then shows that the decomposition in \eqref{eqn:decomposition} remains valid, with finite-dimensional irreducible representations $\rho_i$. In addition, these irreducible representations remain orthogonal and the proof of the lemma goes through as above.

\section{Encoding circuit}
\label{app:circuit}
With the assumptions of Lemma \ref{lem:representation} and a rank-one map $V = |\Phi\rangle \langle \Omega|$, it is possible to obtain an explicit circuit for the encoding map. Note that $|\Omega\rangle$ won't be normalized in general. 

\begin{lemma}
With the assumptions of Lemma \ref{lem:representation}, and a rank-one operator $V$ such that $\tr(V^\dagger V_G) = d$, the encoding map $\mathcal{E}: \mathcal{H}_L \to \mathcal{H}_P$ takes the simple form:
\begin{align}
\mathcal{E}(|\psi\rangle) = U |\psi\rangle |\phi\rangle
\end{align}
where $U$ is a unitary matrix that decomposes $\pi$ as a sum of irreducible representations, and $|\phi\rangle$ is an arbitrary state in the multiplicity space of the irreducible representation $\lambda$.
\end{lemma}

\begin{proof}
Let us write $\lambda = \rho_0$ and decompose the physical representation $\pi$ as a sum of irreducible representations $\rho_k$, with their associated multiplicity space $\M_k$:
\begin{align}
\pi(g) = U \left[ \bigoplus_k \rho_k(g) \otimes \1_{\M_k} \right] U^\dag,
\end{align}
where the intertwiner operator $U$ is assumed to be unitary by Lemma \ref{lem:U}.

The operator $V$ can be written as $V = |\Phi\rangle \langle \Omega|$, with $|\Phi\rangle \in \mathcal{H}_L$ and $|\Omega\rangle \in \mathcal{H}_L$. 
The encoding map is 
\begin{align}
V_G &= \frac{1}{|G|} \sum_{g \in G} \pi(g) |\Phi\rangle \langle \Omega|\lambda(g)^\dagger \\
&= \frac{1}{|G|} \sum_{g \in G} U \left( \bigoplus_k \rho_k(g) \otimes \1_{\M_k}\right)  U^\dagger|\Phi\rangle \langle \Omega|\rho_0(g)^\dagger \\
&= \frac{1}{d } \sum_{p,q=0}^{d-1} U \left( |p\rangle\langle q| \otimes \1_{\M_0}\right)U^\dagger  |\Phi\rangle \langle \Omega|q\rangle \langle p|  & \text{(Schur orthogonality)}\\
&= \frac{1}{d } \sum_{p,q=0}^{d-1} U \left( |p\rangle \langle \Omega|q\rangle \langle q| \otimes \1_{\M_0}\right) U^\dagger |\Phi\rangle \langle p| \\
&= \frac{1}{d } \sum_{p=0}^{d-1} U \left( |p\rangle \langle \Omega|  \otimes \1_{\M_0}\right) U^\dagger |\Phi\rangle \langle p| 
\end{align}
and applied to a state $|\psi\rangle \in \mathcal{H}_L$, it gives
\begin{align}
\mathcal{E}(|\psi\rangle) &= \frac{1}{d  } \sum_{p=0}^{d-1} U \left( |p\rangle \langle \Omega|  \otimes \1_{\M_0}\right) U^\dagger |\Phi\rangle \langle p| \psi\rangle\\
&=\frac{1}{d  } U \left( |\psi \rangle \langle \Omega|  \otimes \1_{\M_0}\right) U^\dagger |\Phi\rangle\\
&= U |\psi\rangle |\phi\rangle
\end{align}
where we defined
\begin{align}
|\phi\rangle = \frac{1}{d}   \langle \Omega  | U^\dagger  |\Phi\rangle. 
\end{align}
Note that this state is normalized since $V_G$ is an isometry. 
\end{proof}

\section{The GKP code}
\label{app:gkp}

The GKP construction is another interesting example that deviates from the setup of Lemma \ref{lem:representation} because the relevant group is infinite. 
Let $G = H_3(\mathbb{Z}) = \langle x,y \rangle$ be the discrete Heisenberg group with 
\begin{align}
x= \begin{bmatrix} 1 & 1 &0 \\ 0 & 1 & 0 \\ 0 & 0 & 1 \end{bmatrix}, \qquad y = \begin{bmatrix} 1 & 0 & 0 \\ 0 & 1 & 1 \\ 0 & 0 & 1\end{bmatrix}.
\end{align}
The two generators satisfy the relations:
\begin{align}
z= xyx^{-1} y^{-1}, \quad xz=zx, \quad yz=zy,
\end{align}
and a general group element is of the form $y^b z^c x^a = \left[ \begin{smallmatrix} 1 & a & c\\ 0 & 1 & b \\ 0 & 0 & 1\end{smallmatrix} \right]$, for arbitrary $a, b, c \in \mathbbm{Z}$. 

We consider two representations of the Heisenberg group. 
The $d$-dimensional logical representation is given by
\begin{align}
\lambda(x) = X, \qquad \lambda(y) = Z,
\end{align}
where $X$ and $Z$ are the $d$-dimensional Pauli matrices which satisfy
\begin{align}
X^d = Z^d = \mathbbm{1}_d, \qquad XZX^{-1} Z^{-1} = \omega^{-1} \1_d,
\end{align}
with $\omega = e^{2\pi i/d}$.
The physical representation $\pi$ is infinite-dimensional and acts as displacements on a single bosonic mode with creation and annihilation operators $\hat{a}^\dagger$ and $\hat{a}$:
\begin{align}
\pi(x) = \hat{D}(\alpha), \qquad \pi(y) = \hat{D}(\beta),
\end{align}
where $\alpha, \beta \in \mathbbm{C}$ satisfy $e^{\beta \alpha^* - \beta^* \alpha} = \omega$ and the displacement operators are defined by
\begin{align}
\hat{D}(\alpha) = e^{\alpha \hat{a}^\dagger - \alpha^* \hat{a}}.
\end{align}
This choice ensures that $\pi(x) \pi(y) \pi(x^{-1}) \pi(y^{-1})= \omega^{-1} \mathbbm{1}$.

The images of a general group element $y^b z^c x^a$ for these two representations are given by
\begin{align}
\lambda(y^b z^c x^a) &= \omega^{-c} Z^b X^a\\
\pi(y^b z^c x^a) &= \omega^{-c} \hat{D}(b \beta) \hat{D}(a \alpha) =  \omega^{-c + ab/2} \hat{D}(a \alpha + b \beta),
\end{align}
where we used that $\hat{D}(\gamma) \hat{D}(\delta) = e^{(\gamma \delta^* - \gamma^* \delta)/2} \hat{D}(\gamma + \delta)$, for any complex numbers $\gamma, \delta$.
Note that for these representations, it suffices to take $c \in \mathbbm{Z}/d\mathbbm{Z}$.
The natural choice of physical state here is the vacuum state: $|\Phi\rangle = |0\rangle$:
\begin{align}
\mathcal{E}(|k\rangle) &\propto \sum_{g \in H_3(\mathbbm{Z})} \pi(g) |0\rangle \langle \Omega | \lambda(g)^\dagger |k\rangle\\
&\propto \sum_{a,b \in \mathbbm{Z}} \sum_{c \in \mathbbm{Z}/d\mathbbm{Z}} \omega^{-c + ab/2} \hat{D}(a \alpha + b \beta) |0\rangle \langle \Omega | \omega^{c} X^{-a} Z^{-b} |k\rangle\\
&\propto d \sum_{a,b \in \mathbbm{Z}} \omega^{ab/2} \hat{D}(a \alpha + b \beta) |0\rangle \langle \Omega |   X^{-a} Z^{-b}|k\rangle\\
&\propto d \sum_{a,b \in \mathbbm{Z}} \omega^{ab/2-bk} \hat{D}(a \alpha + b \beta) |0\rangle \langle \Omega |   X^{-a} |k\rangle\\
&\propto d \sum_{a,b \in \mathbbm{Z}} \omega^{ab/2-bk} | a \alpha + b \beta \rangle \langle \Omega  |k-a \mod d\rangle
\end{align}
Choosing $|\Omega\rangle \propto |0\rangle \in \mathbbm{C}^d$ gives
\begin{align}
\mathcal{E}(|k\rangle)  \propto d \sum_{a,b \in \mathbbm{Z}} \omega^{abd/2 - bk/2} | (da + k) \alpha + b \beta \rangle.
\end{align}
In particular, the choice $d=2$, $\alpha = \sqrt{\pi/2}$, $\beta = i \sqrt{\pi/2}$ yields the standard GKP qubit with a square lattice.

\section{Removing the assumption that $\lambda$ is irreducible in Lemma \ref{lem:representation}}
\label{app:general}

Lemma \ref{lem:representation} only applies when the logical representation $\lambda$ is irreducible. This is not the most general case, and indeed, simple examples such as the bosonic cat codes do not satisfy this assumption. In fact, if the group $G$ is Abelian, all its irreducible representations are 1-dimensional, which means that the logical representation can never be irreducible as soon as the logical space has dimension at least 2.

Let us therefore consider the encoding when both representations are unitary, but without the assumption that the logical representation $\lambda$ is irreducible. As before, there exist unitaries $U_L$ and $U_P$ on the spaces $\mathcal{H}_L$ and $\mathcal{H}_P$ such that, for all $g\in G$,
\begin{align}
\lambda(g) = U_L \left( \bigoplus_i \rho_i(g)_{\L_i} \otimes \1_{\M_i}\right) U_L^\dagger, \qquad \pi(g) = U_P \left(\bigoplus_i \rho_i(g)_{\L_i} \otimes \1_{\N_i}\right)U_P^\dagger,
\end{align}
where $\{\M_i\}$ and $\{\N_i\}$ are the multiplicity spaces associated with the various irreducible representations of $G$ for $\lambda$ and $\pi$. 
For simplicity, we denote by $\L_i$ the space on which $\rho_i(g)$ acts, even though the irreducible representations of $\lambda$ and $\pi$ do not act on the same space, but merely on isomorphic spaces.

The encoding map is obtained by averaging a linear map $V: \mathcal{H}_L \to \mathcal{H}_P$ through
\begin{align*}
V_G = \frac{1}{|G|} \sum_{g\in G} \pi(g) V \lambda(g)^\dagger.
\end{align*}
To decide whether this is an isometry, we compute $V_G^\dagger V_G$:
\begin{align}
V_G^\dagger V_G&= \frac{1}{|G|^2} \sum_{g,h\in G} \lambda(g) V^\dagger \pi(g^{-1} h) V \lambda(h^{-1})\\
&=\frac{1}{|G|^2} \sum_{g,h \in G} \lambda(g) V^\dagger \pi(h) V \lambda(h)^\dagger \lambda(g)^\dagger & \text{(change of variables)} \label{K5}
\end{align}
Focussing first on the sum on $h$, and writing both representations as sums of irreducible representations and applying Schur's orthogonality relations, we get
\begin{align}
\frac{1}{|G|} \sum_{h\in G} V^\dagger \pi(h) V \lambda(h)^\dagger&= \frac{1}{|G|}\sum_{h\in G} V^\dagger U_P \left( \bigoplus_i \rho_i(h) \otimes \1_{\N_i} \right) U_P^\dagger V U_L \left(\bigoplus_j \rho_j(h)^\dagger \otimes \1_{\M_i}\right) U_L^\dagger\\
&= \bigoplus_{i} \frac{1}{d_i} \sum_{p,q=0}^{d_i-1} V^\dagger U_P ( |p,i\rangle \langle q,i| \otimes \1_{\N_i})  U_P^\dagger V U_L (|q,i\rangle \langle p,i| \otimes \1_{\M_i}) U_L^\dagger
\end{align}
where $\{|p,i\rangle\}_{0\leq p < d_i}$ is an orthonormal basis of $\L_i$. 

Injecting this expression into \eqref{K5} and applying again Shur's orthogonality relations yields
\begin{align}
V_G^\dagger V_G&= \frac{1}{|G|}  \bigoplus_{i} \frac{1}{d_i} \sum_{p,q=0}^{d_i-1}  \sum_{g\in G}  \lambda(g) V^\dagger U_P ( |p,i\rangle \langle q,i| \otimes \1_{\N_i})  U_P^\dagger V U_L (|q,i\rangle \langle p,i| \otimes \1_{\M_i}) U_L^\dagger \lambda(g)^\dagger\\
&=  \bigoplus_{i} \frac{1}{d_i} \sum_{p,q=0}^{d_i-1} \bigoplus_j \frac{1}{d_j} \sum_{r,s=0}^{d_j-1} U_L (|r,j\rangle \langle s,j| \otimes \1_{\M_j}) U_L^\dagger V^\dagger U_P \nonumber \\
& \qquad \times  ( |p,i\rangle \langle q,i| \otimes \1_{N_i})  U_P^\dagger V U_L (|q,i\rangle \langle p,i| \otimes \1_{\M_i}) U_L^\dagger U_L (|s,j\rangle \langle r,j| \otimes \1_{\M_j})U_L^\dagger\\
&=  \bigoplus_{i} \frac{1}{d_i^2} \sum_{p,q,r=0}^{d_i-1}  U_L (|r,i\rangle \langle p,i| \otimes \1_{\M_i}) U_L^\dagger V^\dagger U_P   ( |p,i\rangle \langle q,i| \otimes \1_{\N_i})  U_P^\dagger V U_L (|q,i\rangle \langle r,i| \otimes \1_{\M_i}) U_L^\dagger \label{K8}
\end{align}

Let us define the projectors $\Pi_i^\lambda$ and $\Pi_i^\pi$ on the isotypic component of $\rho_i$ of the representations $\lambda$ and $\pi$ onto the irreducible representation $\rho_i$:
\begin{align}
\Pi_i^\lambda := \frac{\mathrm{dim}(\rho_i)} {|G|} \sum_{g\in G} \tr (\rho_i(g))^* \lambda(g), \qquad \Pi_i^\pi:=  \frac{\mathrm{dim}(\rho_i)} {|G|} \sum_{g\in G} \tr (\rho_i(g))^* \pi(g).
\end{align}
These project onto $U_L(\L_i \otimes \M_i)U_L^\dag$ and $U_P(\L_i \otimes \N_i)U_P^\dag$, respectively.

Then one can define the following operator $\widetilde{V}_i : \L_i \otimes \M_i \to \L_i \otimes \N_i$
\begin{align}
\widetilde{V}_i :=U_P^\dagger \Pi_i^\pi  V \Pi_{i}^\lambda U_L.
\end{align}

Observe that $\sum_{q=0}^{d_i-1}( \langle q,i| \otimes \1_{\N_i})  U_P^\dagger V U_L (|q,i\rangle \otimes \1_{\M_i})$ is nothing but the partial trace $\tr_{\L_i} (\widetilde{V}_i)$. 
In particular, \eqref{K8} simplifies to
\begin{align}
V_G^\dagger V_G = U_L \left( \bigoplus_i \1_{\L_i} \otimes \left(\frac{1}{d_i} \tr_{\L_i} (\widetilde{V}_i)  \right)^\dagger \left(\frac{1}{d_i} \tr_{\L_i} (\widetilde{V}_i)  \right) \right) U_L^\dagger,
\end{align}
which shows that $V_G$ is an isometry if and only if each of the maps $\frac{1}{d_i} \tr_{\L_i} (\widetilde{V}_i)$ are isometries from $\M_i$ to $\N_i$. 

\begin{corol} \label{corol}
If the multiplicity spaces of the logical representation are all trivial, \textit{i.e.} $\mathrm{dim} (\M_i)=1$, then for any physical state $|\Phi\rangle \in \mathcal{H}_P$, it is possible to find a (generally unnormalized) vector $|\Omega\rangle \in \mathcal{H}_L$ such that the resulting map $V_G$ associated with $V = |\Phi\rangle \langle \Omega|$ is an isometry.
\end{corol}

\begin{proof}
Taking $V = |\Phi\rangle \langle \Omega|$ gives $\widetilde{V}_i = |\Phi_i\rangle \langle \Omega_i|$ for $|\Phi_i\rangle \in \L_i \otimes \N_i$ and $|\Omega_i\rangle \in \L_i \otimes \M_i \cong \L_i$. 
The partial trace $\frac{1}{d_i} \tr_{\L_i} (\widetilde{V}_i)$ is then a vector of $V_i$, and one can choose the norm of $|\Omega_i\rangle$ to enforce the desired normalization of $\frac{1}{d_i} \tr_{\L_i} (\widetilde{V}_i)$.
\end{proof}

\section{Additional gates for multimode bosonic codes}
\label{app:gates}

We focus here on two codes obtained respectively for the Pauli group $\langle X, Z\rangle$ and the Clifford group $\langle \mathrm{H}, \mathrm{S}\rangle$.
For the Pauli group, we explain how to obtain phase-gates as well as the entangling $CZ$ gate. For the Clifford group, we explain how to use quartic Hamiltonians to get the $CZ$ and $T$-gates. 

We denote by $|\overline{k}\rangle$ the encoded state $\mathcal{E}(|k\rangle)$.

\subsection{Pauli code $\langle X,Z\rangle$}
\label{app:pauli}

In this subsection, we consider the standard $S$ gate given by $S = \left[ \begin{smallmatrix} 1 & 0 \\ 0 & i\end{smallmatrix}\right]$.

\begin{lemma}\label{lem:pauli-quartic}
For the Pauli code associated to the group $\langle X, Z\rangle$, the single-qubit logical phase gate $\overline{S} = |\overline{0}\rangle \langle \overline{0}| + i |\overline{1}\rangle \langle \overline{1}|$ and the two-qubit logical controlled-$Z$ gate $\overline{CZ}$ are obtained by quartic Hamiltonians:
\begin{align}
i^{\hat{n}_2^2} = \overline{S}, \qquad (-1)^{\hat{n}_2 \hat{n}_4} = \overline{CZ}.
\end{align} 
\end{lemma}

\begin{proof}
By construction of the code, the logical $Z$ operator, denoted $\overline{Z}$, is obtained as $\overline{Z} = \rho(Z) = (-1)^{\hat{n}_2}$, and therefore $(-1)^{\hat{n}_2} |\overline{k}\rangle = (-1)^k |\overline{k}\rangle$ for $k \in \{0,1\}$. 

It is easy to check that for any integer $n$, it holds that
\begin{align} \label{eqn:function}
 i^{n^2} = e^{i\frac{\pi}{4}} e^{-i \frac{\pi}{4} (-1)^n},
\end{align}
which immediately implies that
\[ i^{\hat{n}_2^2} |\overline{k} \rangle = e^{i\frac{\pi}{4}} e^{-i \frac{\pi}{4} (-1)^{\hat{n}_2}}|\overline{k}\rangle =e^{i\frac{\pi}{4}} e^{-i \frac{\pi}{4} (-1)^{k}}|\overline{k}\rangle= i^{k^2} |\overline{k}\rangle = i^k |\overline{k}\rangle,\]
showing that $i^{\hat{n}_2^2}$ implements a logical $S$ gate. 

Similarly, for any pair of integers $m,n$, it holds that
\[ (-1)^{mn} = e^{i \frac{\pi}{4} (1-(-1)^m )(1-(-1)^n)}, \]
and therefore
\begin{align*}
(-1)^{\hat{n}_2 \hat{n}_4} |\overline{k}\rangle |\overline{\ell}\rangle  &= e^{i \frac{\pi}{4} (1-(-1)^{\hat{n}_2} )(1-(-1)^{\hat{n}_4})}|\overline{k}\rangle |\overline{\ell}\rangle \\
&= e^{i \frac{\pi}{4}(1-(-1)^k)(1-(-1)^\ell)} |\overline{k}\rangle|\overline{\ell}\rangle\\
&= (-1)^{k \ell} |\overline{k}\rangle|\overline{\ell}\rangle
\end{align*}
which concludes the proof.
\end{proof}

\subsection{Clifford code $\langle \mathrm{H}, \mathrm{S}\rangle$}
\label{app:proof3}
For the Clifford code, we recall that we consider variants of the Hadamard and $S$ gates that belong to $SU(2)$:
\[ \mathrm{H} := \frac{1}{\sqrt{2}} \begin{bmatrix} \eta & \eta \\ -\eta^{-1} & \eta^{-1}\end{bmatrix}, \quad \mathrm{S} := \begin{bmatrix} \eta & 0 \\ 0 & \eta^{-1} \end{bmatrix}.\]

\begin{lemma}\label{lem:cliff-quartic}
For the Clifford code associated to the group $\langle \mathrm{H}, \mathrm{S}\rangle$, the single-qubit logical $T$-gate $\overline{T} = |\overline{0}\rangle \langle \overline{0}| + e^{i \pi/4} |\overline{1}\rangle \langle \overline{1}|$ and the two-qubit logical controlled-$Z$ gate $\overline{CZ}$ are obtained by Hamiltonians:
\begin{align}
e^{i \frac{\pi}{16} (\hat{n}_1-\hat{n}_2-1)^2} = \overline{T}, \qquad e^{i \frac{\pi}{4} (\hat{n}_1 - \hat{n}_2 -1)(\hat{n}_3 - \hat{n}_4 -1) }  = \overline{CZ}.
\end{align} 
\end{lemma}

\begin{proof}

The property that the logical $\mathrm{S}$ operator can be implemented with the Gaussian unitary $\rho(\mathrm{S})$ gives 
\[ e^{i\frac{\pi}{4} (\hat{n}_1 - \hat{n}_2)} |\overline{k}\rangle = e^{i \frac{\pi}{4} (1-2k)}|\overline{k}\rangle\]
and therefore
\begin{align}\label{eqn:cond1}
e^{i\frac{\pi}{4} (\hat{n}_1 - \hat{n}_2-1)} |\overline{k}\rangle = (-i)^k|\overline{k}\rangle
\end{align}
for $k\in \{0,1\}$.

We remark that the operator $e^{i\frac{\pi}{4} (\hat{n}_1 - \hat{n}_2-1)}$ can be written as 
\begin{align}
e^{i\frac{\pi}{4} (\hat{n}_1 - \hat{n}_2-1)} = \sum_{\ell = 0}^7 e^{i \frac{\pi}{4} \ell} \, \Pi_\ell,
\end{align}
where 
\[ \Pi_\ell = \sum_{\substack{ n_1, n_2 \, \text{s.t.} \\ n_1-n_2-1 \equiv \ell \mod 8} } |n_1 \rangle \langle n_1| \otimes |n_2\rangle \langle n_2| \]
is a projector on the space spanned by Fock states $|n_1\rangle |n_2\rangle$ satisfying $n_1-n_2-1 \equiv \ell \mod 8$.

From \eqref{eqn:cond1}, one can infer that 
\[ \Pi_0 |\overline{0}\rangle = |\overline{0}\rangle, \qquad \Pi_{-2} |\overline{1}\rangle = |\overline{1}\rangle.\]
We want to understand how $e^{i \frac{\pi}{16} (\hat{n}_1-\hat{n}_2-1)^2}$ acts on the code space. In particular, it is immediate that if $n_1-n_2-1 \equiv 0 \mod 8$, then $(n_1-n_2-1)^2 \equiv 0 \mod 64$ and if $n_1-n_2-1\equiv -2 \mod 8$, then $(n_1-n_2-1)^2 \equiv 4 \mod 32$. This shows that the operator $e^{i \frac{\pi}{16} (\hat{n}_1-\hat{n}_2-1)^2}$ acts trivially on the support of $\Pi_0$ and acts like $e^{i \frac{\pi}{16} 4} = e^{i\frac{\pi}{4}}$ on the support of $\Pi_{-2}$. In other words,
\begin{align*}
e^{i \frac{\pi}{16} (\hat{n}_1-\hat{n}_2-1)^2} |\overline{k}\rangle &= e^{i \frac{\pi}{4} k} |\overline{k}\rangle,
\end{align*}
which shows that it implements a logical $T$ gate. 

Similarly, for $k, \ell \in \{0,1\}$, if $n_1-n_2-1 \equiv -2k \mod 8$ and $n_3-n_4-1 \equiv -2\ell \mod 8$, then 
\[ (n_1-n_2-1)(n_3-n_4-1) \equiv 4 k \ell \mod 16\]
and therefore
\[ e^{i \frac{\pi}{4} (\hat{n}_1 - \hat{n}_2 -1)(\hat{n}_3 - \hat{n}_4 -1) } \Pi_{-2k} \otimes \Pi_{-2\ell} = (-1)^{k\ell} \Pi_{-2k} \otimes \Pi_{-2\ell},\]
which shows that
\[ e^{i \frac{\pi}{4} (\hat{n}_1 - \hat{n}_2 -1)(\hat{n}_3 - \hat{n}_4 -1) } |\overline{k}\rangle |\overline{\ell}\rangle= (-1)^{k\ell} |\overline{k}\rangle |\overline{\ell}\rangle. \qedhere\]
\end{proof}

\section{Knill-Laflamme conditions for the pure-loss channel}
\label{app:KL}

We study here the effect of the pure-loss channel for the bosonic Pauli code defined from the group $\langle X,Z\rangle$, but similar considerations also apply to other variants of the Pauli code or of the Clifford code. 

The Kraus operators of the two-mode pure-loss channel $\mathcal{L}_\gamma$ with loss rate $\gamma \in [0,1)$ are given by~\cite{AND18}:
\begin{align*}
E_{p_1,p_2} = \left(\frac{\gamma}{1-\gamma}\right)^{(p_1+p_2)/2} \frac{\hat{a}_1^{p_1} \hat{a}_2^{p_2}}{\sqrt{p_1! p_2!}} (1-\gamma)^{(\hat{n}_1+\hat{n}_2)/2},
\end{align*}
so that the action of the channel $\mathcal{L}_\gamma$ on an arbitrary two-mode state $\rho$ is 
\begin{align}\label{eqn:L}
\mathcal{L}_\gamma(\rho)  = \sum_{p_1, p_2=0}^\infty E_{p_1, p_2} \rho E_{p_1,p_2}^\dag.
\end{align}

A straightforward calculation shows that these Kraus operators attenuate coherent states: defining $\mu := \sqrt{1-\gamma}$, we get
\[\hat{a}^p (1-\gamma)^{\hat{n}/2} |\alpha\rangle = \left( \mu \alpha\right)^p e^{-\gamma |\alpha|^2/2} |\mu \alpha\rangle.\]

We can apply $E_{p_1,p_2}$ to a product of two cat states $|c_j (\alpha_j)\rangle |c_{1-j}(\alpha_{1-j})\rangle$ for $j \in \{0,1\}$, it gives
\begin{align*}
E_{p_1,p_2}|c_j (\alpha_j)\rangle |c_{j-1}(\alpha_{j-1})\rangle &=  \left(\frac{\gamma}{1-\gamma}\right)^{(p_1+p_2)/2} \frac{(\mu \alpha_j)^{p_1} (\mu \alpha_{1-j})^{p_2}}{\sqrt{p_1! p_2!}} e^{-\gamma (|\alpha_j|^2 + |\alpha_{1-j}|^2)/2} |c_{j-p_1} (\mu \alpha_j)\rangle |c_{1-j-p_2}(\mu \alpha_{1-j})\rangle\\
&=  e^{-\gamma (|\alpha_j|^2 + |\alpha_{1-j}|^2)/2} \frac{ \gamma^{(p_1+p_2)/2}  \alpha_j^{p_1} \alpha_{1-j}^{p_2}}{\sqrt{p_1! p_2!}}  |c_{j-p_1} (\mu \alpha_j)\rangle |c_{1-j-p_2}(\mu \alpha_{1-j})\rangle\\
&= f(\alpha_j, \alpha_{1-j}, p_1, p_2)  |c_{j-p_1} (\mu \alpha_j)\rangle |c_{1-j-p_2}(\mu \alpha_{1-j})\rangle
 \end{align*}
with indices taken modulo 2, and we defined the function $f(\alpha, \beta, p,q) := e^{-\gamma( |\alpha|^2+|\beta|^2)/2} (\alpha \sqrt{\gamma})^p (\beta\sqrt{\gamma})^q/\sqrt{p!q!}$.

Focusing on the Pauli code with initial state $|\alpha\rangle |\alpha e^{i \theta}\rangle$, and denoting $|\overline{k}\rangle := \mathcal{E}(|k\rangle)$, one can check the Knill-Laflamme conditions for the Kraus operators of the pure-loss channel by applying the previous expression for $\alpha_0 := \alpha e^{i \theta}, \alpha_1 := \alpha$, 
\begin{align*}
\langle \overline{k} | E_{p_1, p_2}^\dag E_{q_1, q_2} |\overline{\ell}\rangle &\propto  f(\alpha_k, \alpha_{1-k}, p_1, p_2)^* f(\alpha_\ell, \alpha_{1-\ell}, q_1, q_2) \langle c_{k-p_1} (\mu \alpha_k)|c_{\ell-q_1} (\mu \alpha_\ell)\rangle \langle c_{1-k-p_2}(\mu \alpha_{1-k})|c_{1-\ell-q_2}(\mu \alpha_{1-\ell})\rangle. 
\end{align*}
Diagonal terms of the form $\langle \overline{k} | E_{p_1, p_2}^\dag E_{q_1, q_2} |\overline{k}\rangle$ are nonzero only if $p_1=q_1$ and $p_2=q_2$ since the even and odd cat states are orthogonal (they have support on the even and odd Fock states, respectively). In that case, we have
\begin{align*}
\langle \overline{k} | E_{p_1, p_2}^\dag E_{p_1, p_2} |\overline{k}\rangle &\propto   f(\alpha_k, \alpha_{1-k}, p_1, p_2)^* f(\alpha_k, \alpha_{1-k}, p_1, p_2)  \\
&=  e^{-\gamma (|\alpha_k|^2 + |\alpha_{1-k}|^2)} \frac{ \gamma^{(p_1+p_2)}  |\alpha_k|^{2p_1} |\alpha_{1-k}|^{2p_2}}{{p_1! p_2!}} 
\end{align*}
which is independent of $k$, provided that $|\alpha_0| = |\alpha_1|$. 

Moreover, for nondiagonal terms, we observe that if $\theta$ is not a multiple of $\pi$, then the overlaps between the cat states of amplitude $\mu \alpha_0$ and $\mu \alpha_1$ always vanish in the limit of large energy, $\alpha \to \infty$,
\[ \lim_{\alpha \to \infty} \langle \overline{0} | E_{p_1, p_2}^\dag E_{q_1, q_2} |\overline{1}\rangle =0.
\]

This suggests to optimize the choice of the phase $\theta$ so as to maximize the distance between the constellations of coherent states for the encoded states $|\overline{0}\rangle$ and $|\overline{1}\rangle$. A similar observation was made in \cite{JIB23} where this distance was computed explicitly for many families of quantum spherical codes.
In the case of the Pauli code, we see that the choices $\theta = 0$ and $\theta=\pi/2$ are respectively the worst and best choices with that respect. This is confirmed numerically, and can also be seen on Fig.~2 in the main text.

\section{Expression of the Clifford code for an arbitrary initial state}
\label{app:cliff}

There are various representations of the Clifford group. We consider here the binary octahedral group $2O$, with presentation $\langle s,t | (st)^2 = s^3=t^4\rangle$ and generators $s = \frac{1}{2}(1+i+j+k)$ and $t = \frac{1}{\sqrt{2}}(1+i)$. It has the advantage of containing only 48 elements. 
Using the standard representation of a quaternion $a+bi+cj+dk$ as a $2\times 2$ complex matrix $\left[ \begin{smallmatrix} a+bi & c+di\\-c+di & a-bi\end{smallmatrix} \right]$, we obtain the following generators for the group $2O$:
\[ \mathrm{H} := \frac{1}{\sqrt{2}} \begin{bmatrix} \eta & \eta \\ -\eta^{-1} & \eta^{-1}\end{bmatrix}, \quad \mathrm{S} := \begin{bmatrix} \eta & 0 \\ 0 & \eta^{-1} \end{bmatrix},\]
where we defined $\eta = \frac{1}{\sqrt{2}} (1+i)$ and note that $\mathrm{H}$ and $\mathrm{S}$ differ from the standard form of the Hadamard and phase gates because we focus here on operators in $SU(2)$. 
One can readily check that $(\mathrm{H}\mathrm{S})^2 = \mathrm{H}^3= \mathrm{S}^4=-\1_2$.

The 48 matrices composing the group consist of 8 diagonal matrices, 8 antidiagonal matrices and 32 Hadamard-like matrices:
\begin{align*}
&\begin{bmatrix} \eta^k & 0 \\ 0 & \eta^{-k}\end{bmatrix}, \qquad \begin{bmatrix} 0 & -\eta^{k} \\ \eta^{-k} & 0\end{bmatrix}, \qquad  k \in \{0, \ldots, 7\}, \\
& \frac{1}{\sqrt{2}}\begin{bmatrix} \eta^{2\ell} & \eta^{2m} \\  -\eta^{-2m} & \eta^{-2\ell}\end{bmatrix}, \qquad \frac{1}{\sqrt{2}}\begin{bmatrix} \eta^{2\ell+1} & \eta^{2m+1} \\  -\eta^{-2m-1} & \eta^{-2\ell-1}\end{bmatrix}, \qquad\ell, m \in \{0, 1, 2,3\}.
\end{align*}
One can apply the construction of Lemma \ref{lem:representation} for an arbitrary initial coherent state $|\alpha\rangle |\beta\rangle$ and $|\Omega\rangle \propto |0\rangle$. 
This gives 
\begin{align*}
\mathcal{E}(|0\rangle) \propto \sum_{k=0}^7 \eta^{-k} |\eta^k \alpha\rangle |\eta^{-k} \beta\rangle + \frac{1}{\sqrt{2}} \sum_{\ell,m=0}^3\sum_{p=0}^1 \eta^{-2\ell-p} |(\eta^{2\ell+p} \alpha + \eta^{2m+p} \beta)/\sqrt{2}\rangle |(-\eta^{-2m-p}\alpha + \eta^{-2\ell-p}\beta)/\sqrt{2}\rangle,
\end{align*}
and the state $\mathcal{E}(|1\rangle)$ is obtained by swapping the two modes. 
\end{document}